\documentclass[amsmath,amssymb]{revtex4-1}

\usepackage{babel}
\usepackage{amsmath}
\usepackage{graphicx}
\usepackage{xcolor}
\usepackage[displaymath]{lineno}

\usepackage{varioref}
\usepackage{hyperref}
\usepackage[noabbrev]{cleveref}

\graphicspath{ {./images/} }

\begin{document}
\title{A photonic complex perceptron for ultrafast data processing}
\author{Mattia Mancinelli*, Davide Bazzanella, Paolo Bettotti, Lorenzo Pavesi}

\affiliation{Nanoscience Laboratory, Department of Physics, University of Trento,\\Via Sommarive 14, 38123, Trento, Italy }

\date{\today}

\begin{abstract}
In photonic neural network a key building block is the perceptron.
Here, we describe and demonstrate a complex-valued photonic perceptron that combines time and space multiplexing in a fully passive silicon photonics integrated circuit.
An input time dependent bit sequence is broadcasted into a few delay lines where the relative phases are trained by particle swarm algorithms toward the given task.
Since only the phases of the propagating optical modes are trained, signal attenuation in the perceptron due to amplitude modulation is avoided.
The perceptron performs binary pattern recognition and few bit delayed XOR operations up to 16 Gbps (limited by the used electronics) with Bit Error Rates as low as $10^{-6}$.
The perceptron is fully integrated, silicon based, scalable, and can be used as a building block in large neural networks.
\end{abstract}

\maketitle

\section{Introduction}
Photonic Neural Networks (PNN) are radically changing the benchmark of complexity and speed of computation \cite{xu202111,feldmann2021parallel}.
Photonic Integrated Circuits (PICs) can be used in \textit{beyond von-Neumann} architecture to perform logical operations more complex than the boolean primitives (e.g., speech\cite{vandoorne2011parallel} and image recognition\cite{xu202111}, signal recovery\cite{katumba2019neuromorphic,argyris2018photonic}, object classification \cite{harris2018linear}).
For such complex operations the advantages of photonics (multiwavelength, speed, low power) overcomes its limitations (mainly system complexity, power losses and footprint).
An in depth description of the basics of a PNN can be found in recent review papers \cite{peng2018neuromorphic,genty2020machine}.
Generally, PNNs can be classified into two main categories: feed forward (FFN) and reservoir (RCN) networks.
The former mostly used in computer science as the basis of deep learning algorithms allows the optimization of the network (in terms of weights and biases) through a deterministic algorithm (e.g., the gradient descent or the back-propagation).
Despite their optimized structure, FFNs are not designed to work with time dependent signals and are hardly implemented in high speed signal processing (e.g., to correct high bandwidth optical signals distorted by nonlinear propagation effects or to analyze correlation between signal distant in time).
Several papers demonstrated optical implementation of FFN in PIC as optical accelerator \cite{hughes2018training,harris2018linear}, since matrix multiplication is a fundamental operation in FFNs which fully exploits the benefits of optics: high speed, low power consumption and inherent parallelism .
More recently, innovative approaches that make use of frequency combs to realize deep and convolutional PNN have been reported \cite{feldmann2021parallel,xu2020photonic}.

For the analysis of time dependent signals, recurrence is a nearly mandatory property of the network, as it enriches the network description and unveils (nonlinear) relations between retarded signals (bits) by the network memory.
Recurrence greatly increases the complexity of the network dynamics and prevents the possibility of a detailed description of the instantaneous network state.
Yet, it is heuristically demonstrated that if the reservoir computer (RC) is forced to work "at the edge of chaos", then the network is able to effectively compute complex tasks \cite{carroll2020reservoir}.
On this line, RC is the network paradigm investigated since the first experimental implementation of PNN \cite{appeltant2011information}.
Several papers demonstrated photonics RC made of either a single \cite{duport2012all} or multiple nodes \cite{vandoorne2014experimental} or using passive as well as active PNN \cite{vandoorne2011parallel,tait2014broadcast}.
While RC highly relaxes the requirements of the training phase as its internal state is not trained and the training is a simple linear projection from a (sub)set of the RC node states, they have two main drawbacks when photonic implementation is considered: 
\begin{itemize}
\item since the training is based on heuristic methods, neither the network topology nor the strength of the connections are optimized.
This heavily limits the complete exploitation of the PNN hardware; 
\item spatially distinct RC nodes suffer from propagation losses and coupling losses.
Therefore, the RC recurrence is only partially exploited and the scalability limited.
\end{itemize}

For both kinds of PNNs a key element is the perceptron \cite{rosenblatt1957perceptron}.
During the training, this component weights the input signals to draw a linear decision boundary and achieve the desired task.
Here, we present a complex-valued photonic perceptron that combines time and space multiplexing to acts as a non-linear classifier.
We use the complex nature of the propagating optical mode to represent the inner perceptron state.
We demonstrate a fully passive optical perceptron where only the phases of the modes are learned which avoids additional power loss due to amplitude modulation.
The perceptron makes optical computation at ultrafast speed and solves several logic tasks.
The analysis shows that our perceptron can perform as a basic building block for large scale PNNs.\\
Basically, the perceptron processes time dependent input bit sequences by broadcasting them into a small number of delay lines (waveguides, WG).
The perceptron is trained by optimizing the values of the relative phase of the signal in the various WGs while their amplitudes are fixed by the delay lines losses.
The perceptron perform pattern recognition and delayed XOR task up to 16 Gbps (limited by the testing electronics) with Bit Error Rates (BERs) as low as $10^{-6}$ (that is the statistical limit of the sequences provided to the PNN).
Our perceptron is fully integrated and compatible with CMOS technology, and all its parameters are trained, by using a particle swarm algorithm \cite{part_swarm}, which assures an optimized use of the resources.

\section{The complex perceptron}
\begin{figure}
	\centering
	\includegraphics[]{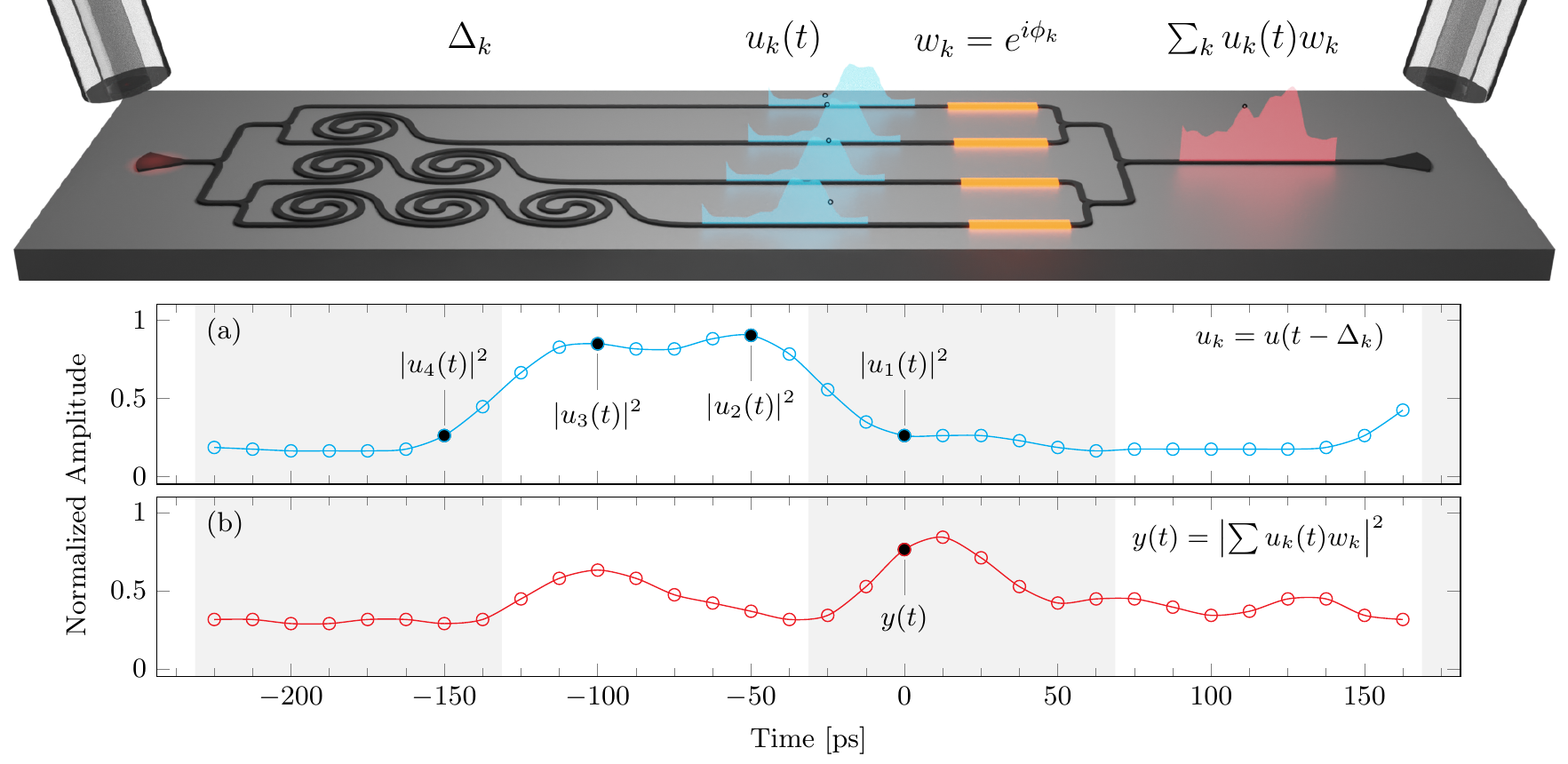}
	\caption{ 
	    (Top) Sketch of the integrated photonic circuit which performs as a complex perceptron.
	    Gratings are used to couple the light in and out of the chip.
	    The red spot on the left shows the input signal.
	    A $1\times4$ splitter distributes the input signal to four delay lines, realized by spirals (each spiral adds a delay of $\Delta _t$).
	    Then, thermal phase modulators (yellow segments) allow controlling the relative phases ($w_k = e^{i\phi_k}$) of the signals ($u_k(t)$, blue lineshapes).
	    These are then summed by combiners and the resulting signal ($y(t)$, red lineshape) is output via a grating.
	    (Bottom) (a) Input signal versus time.
	    The black circles show the sampling of the input signal done by the four delay lines.
	    (b) Perceptron output versus time.
	    This signal is the one that results after the photodetector.
	    The dot shows the time at which the complex sum is performed.}
	\label{fig:dut}
\end{figure}
The integrated version of the complex perceptron is schematically shown in \Cref{fig:dut}.
It is composed by an input grating coupler for TE (Transverse Electric, in plane) polarization connected to a $1\times4$ beam splitter composed of cascaded $1\times2$ multimode interferometers (MMIs).
Therefore, the input signal $u(t)$ is split in four copies ($u_k(t)$, k=1,$\dots$,4 ) that propagate in the four WGs.
The waveguides are spiraled to realize delay lines and, thus, to retard copies of the input signal by an integer multiple of $\Delta _t\ =\ 50\ ps$.
Thus, in each WG (k=1,$\dots$,4) a delayed signal $u_k(t)=u(t-(k-1)\Delta_t)$ is propagating.
Thermal heaters (yellow lines) above the WGs to heat them and, therefore, impart a given phase $\phi_k$ to the delayed input signals.
In this way, phase encoded weights, ${w_k}=\exp(i \phi_k)$, are attributed to each delayed copy.
Finally, the weighted delayed copies $w_ku_k$ are coherently summed by a $4\times1$ combiner realized by cascaded $2\times1$ MMI.
At the output, the signal $\sum_{k=1}^{4} w_k u_k(t)$ is collected by a fiber and detected by a photodetector.
At the detector, the non-linear transformation ($|\cdot|^{2}$) is performed on the output signal.
The time dependent electrical signal $y(t)=|\sum_{k=1}^{4} w_k u_k(t)|^2$ is the output of the perceptron, i.e., the perceptron prediction.
In a nut-shell, the effect of the delay lines is to populate the perceptron input layer by mixing information coming from adjacent bits and the complex perceptron performs logical operations by modulating the interference between different bits, via the phase controls.

\begin{figure}
	\centering
	\includegraphics[scale=1]{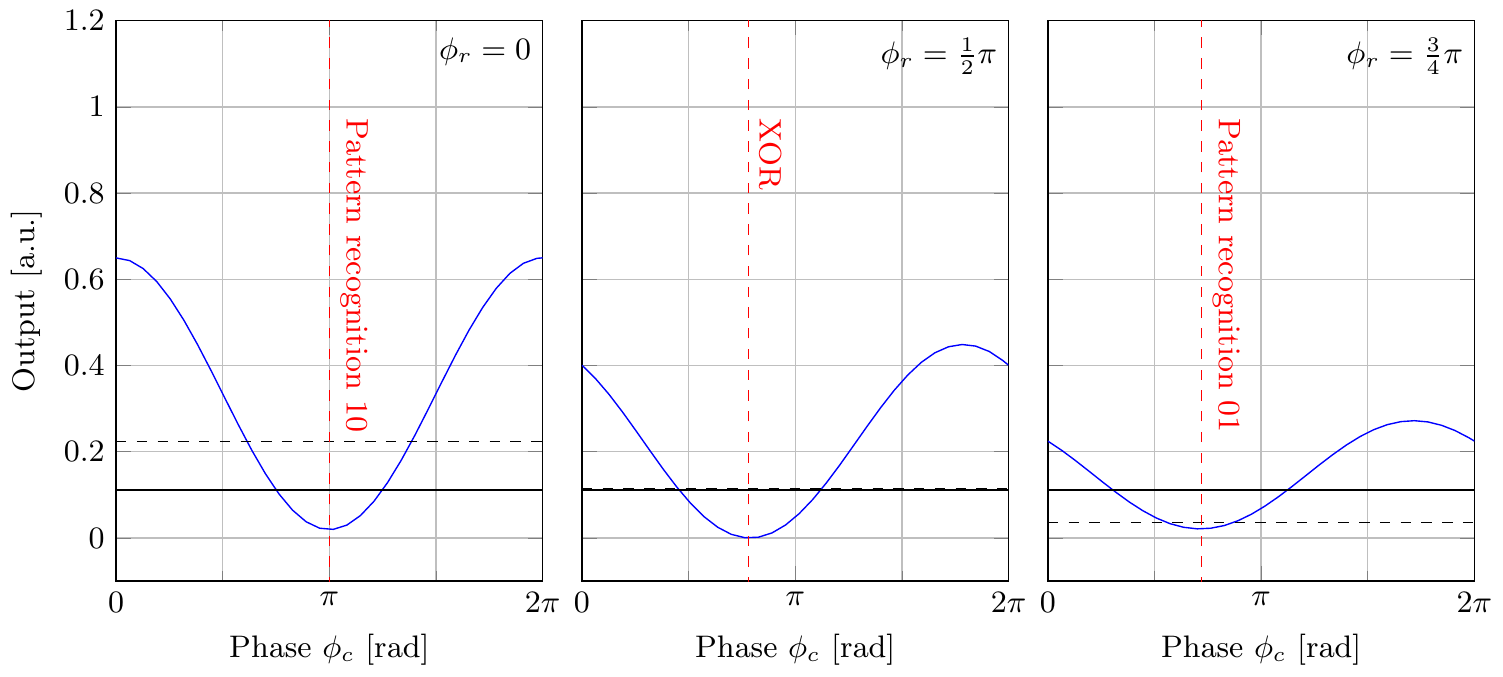}
	\caption{
	    The perceptron prediction for several input states as a function of the relative phase $\phi_c$ between inputs $u_1$ and $u_2, u_3$ for three values of the relative phase $\phi_r$ between $u_2$ and $u_3$.
	    Red dashed lines indicate the optimal common phase required to solve the particular task indicated on the lines.
	    The values of $\{u_1, u_2, u_3\}$ are respectively: $1, 1, 1$ (blue line), $1, 0, 0$ (dashed horizontal line), and $ 0, 1, 1 $ (continuous horizontal line).
    }
	\label{fig:delayed_example}
\end{figure}

To see the action of the phase control, let us consider three signals $u_1,\ u_2,\ u_3$ after the delay stages which have been sampled from a binary sequence.
Two signals are sampled from the same input bit (e.g., $u_2 = u_3$).
This might happen whenever $(\Delta _t)^{-1}$ is faster than the input bit-rate.
If we simplify the discussion by assuming a zero phase delay on the first delay line, then the perceptron output, $y$, is\\
\begin{equation}
		y = |u_1 w_1 + u_2 w_2 + u_3 w_3|^2
\end{equation}
where $w_1= a_{1} e^{\phi_{1}} =1;\ w_{2} = a_{2} e^{\phi_{2}}$ and $\ w_{3} = a_{3} e^{\phi_{3}}$.
The parameter $a_k$ accounts for the different delay line losses.
If we introduce, $\phi_c = \phi_2$ and $\phi_r = \phi_3 - \phi_2$ as the common and the relative phases of the other two signals and $\gamma = a_{3}/a_{2}$.
Then
\begin{equation}
	y = |u_1 + a_2 u_2 e^{i\phi_c} ( 1 + \gamma e^{i \phi_r })|^2\\
\end{equation}

\noindent Assuming a constant $\phi_r$, we can simplify further by introducing a complex valued constant $\eta= a_2 ( 1 + \gamma e^{i \phi_r})$

\begin{equation}
		y = |u_1 + \eta u_2  e^{i\phi_c}|^2 \label{eq:simple}
\end{equation}

\noindent that is the prediction given by the interference between $u_1$  and $u_2$.
Here, it is also highlighted the role of $\phi_r$ which controls the complex amplitude of the interfering signal.
Therefore, despite only the phases of the weights applied to $u_k$ are controlled, the interplay between the different phase delays among the various signals induces a rich combination of the delayed signals in the complex perceptron output.

As an example of use of the complex perceptron, we simulate the simple model of \Cref{eq:simple} and apply it to two binary tasks: two bits pattern recognition and XOR task.
Results are shown in \Cref{fig:delayed_example}, where the various panels report the output signal $y$ as a function of $\phi_c$ for three different $\phi_r$ values.
The three possible bit combinations, e.g., (10), (01), and (11), are represented by the black dotted, black continuous, and blue continuous line, respectively.

The trivial case of both bits zero produces identically null output and it is not shown.
The blue line shows that $y$ results from the sinusoidal interference among the bits when they both are different from zero.
By choosing the proper values for $\phi_c$ and $\phi_r$, the complex perceptron is able to solve both the XOR and the non-trivial pattern recognition task.
The red, vertical dotted lines identify the phase values where the prediction solve the various tasks.
Let us note the important role of $\phi_r$.: \Cref{fig:delayed_example} highlights that $\phi_r$ modulates the amplitude of the interference and the level of the black dashed line.
When $\phi_r = 0$, the system is only able to solve the XOR task with unbalanced high levels and to find the pattern $10$.

\section{Results}
The complex perceptron device creates $N=4$ delayed input copies $u_k(t) = \{ u(t), \ldots, u(t - (k-1) \cdot \Delta_t), \ldots, u(t - (N-1)\cdot \Delta_t )\} $ that populates the feature time series $x_k$.
The perceptron applies the complex weights $w_k = \{ a_1 e^{i \phi_1 }, ... , {a_N e^{i \phi_N }} \}$ and performs the sum and non-linear transformation 
\begin{equation}
	y(t) = |\sum_{k=1}^N u_k(t) w_k |^2  \label{Eq:sum}
\end{equation}
to produce the predictor $y(t)$.
In the actual implementation, amplitudes are fixed by the propagation losses in the spirals and only the phases are trained.
The nominal differential delay and amplitudes are $\Delta_t = 50\ ps$ and $a^2_k = \{1, 0.58, 0.34, 0.2\}$.
Dispersive effects in the spirals are negligible since the dispersion length $L_D = T_0^2/|\beta_2|$ for a 20 Gbps signal is estimated more than 780 m.
A memory of 3 bits, for a binary input signal, is reached at the design bit-rate of 20\ Gbps.
The output at any given time is the coherent weighted sum of the four delayed copies of the input, non-linearly transformed by the detector.

\subsection{Pattern recognition task}
\begin{figure}
	\centering
	\includegraphics[scale=1]{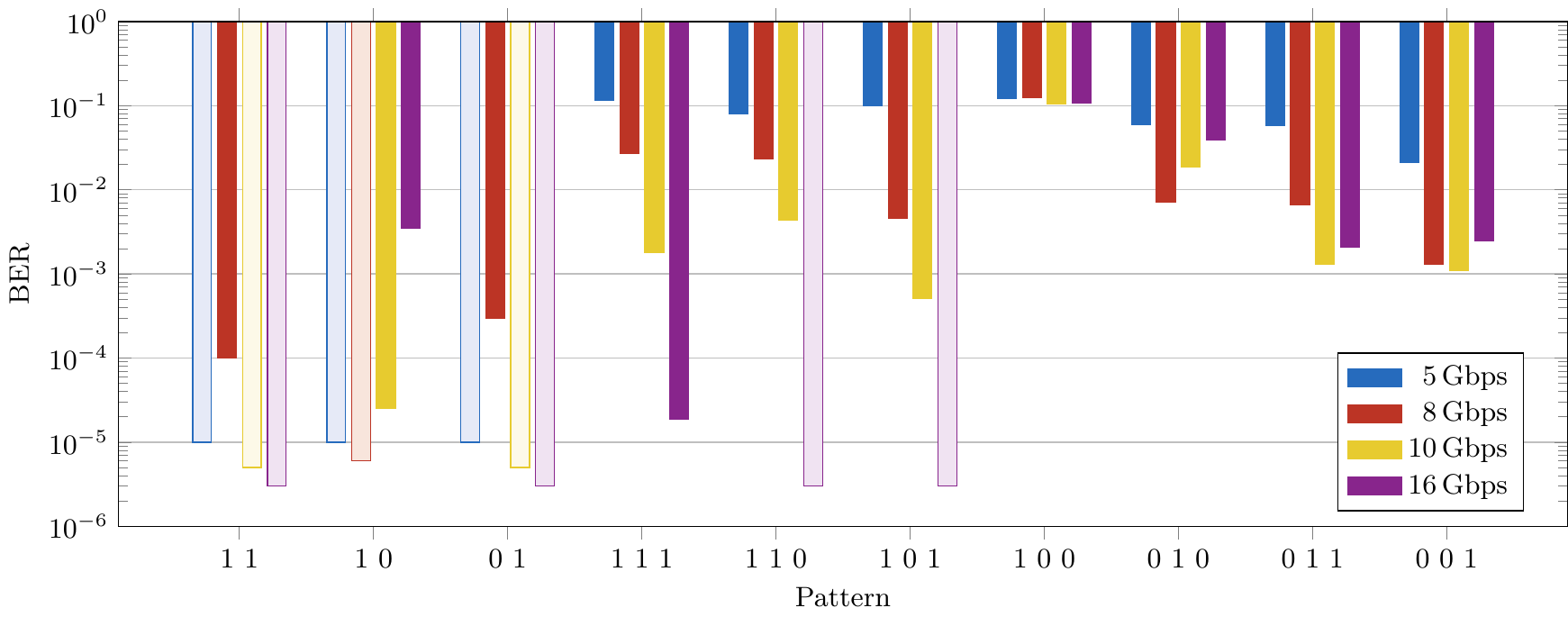}
	\caption{
	    Measured BER as a function of the target symbol for the pattern recognition task.
	    The different colored bars refer to the different bit-rates reported in the inset.
	    Transparent bars refer to a BER value equal to the statistical limit (i.e., error free calculation).
	    This value is $10^{-5}, 6 \times 10^{-6}, 5\times 10^{-6}, 3 \times 10^{-6}$ for a bit-rate of 5, 8, 10, 16 Gbps, respectively.
    }
	\label{fig:pattern_vs_bitrate}
\end{figure}
\begin{figure}
	\centering
	\includegraphics[scale=1]{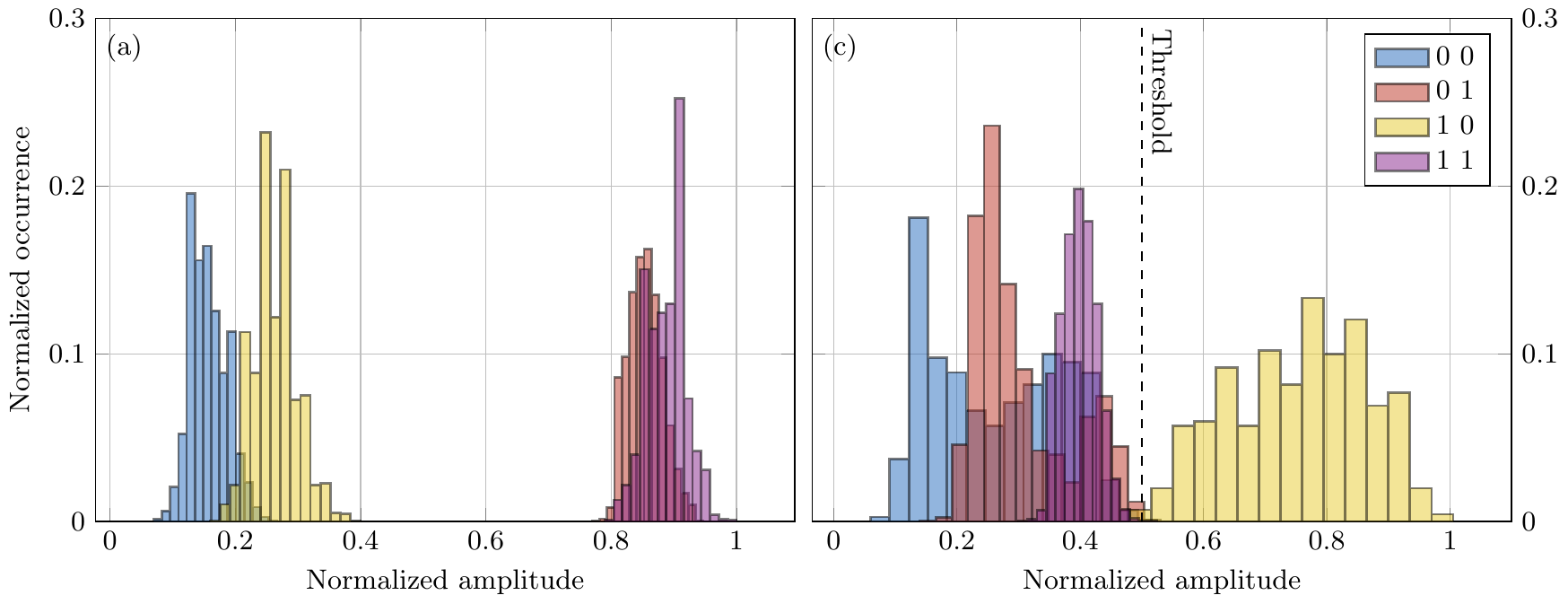}
	\\
	\includegraphics[scale=1]{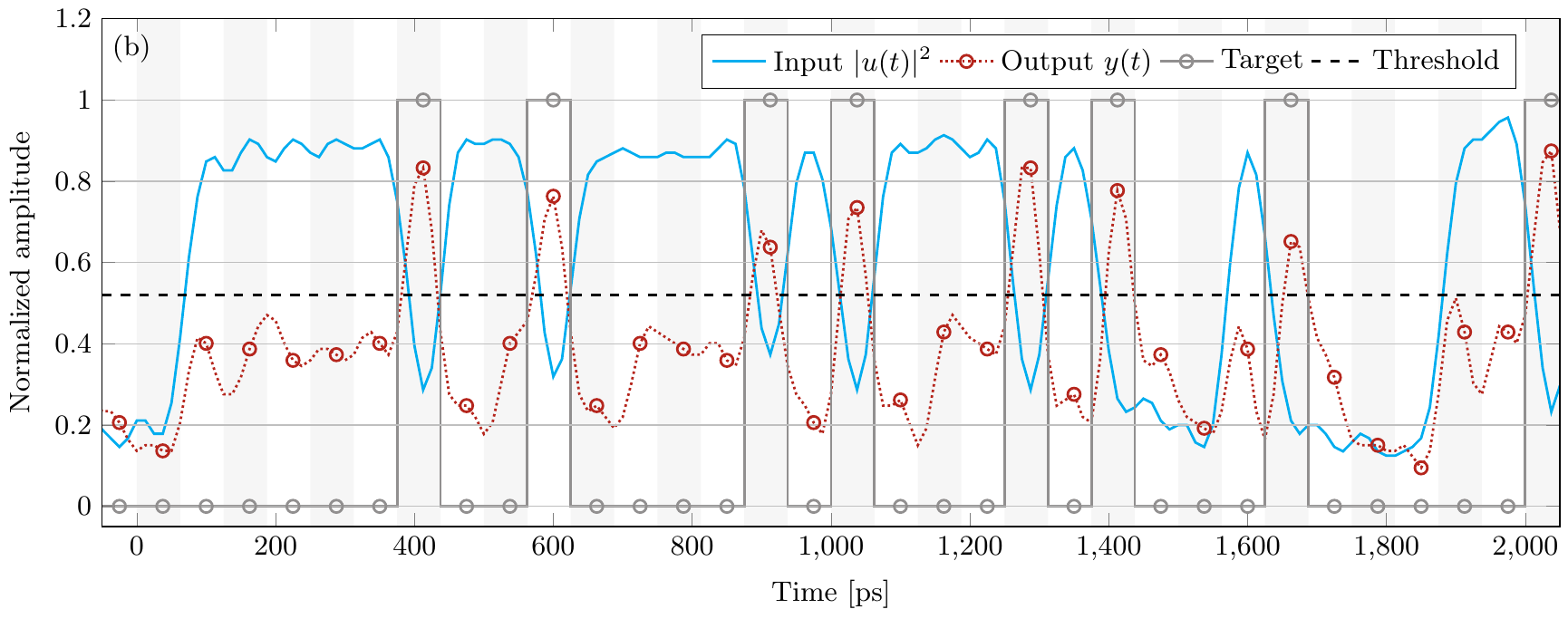}
	\caption{
	    Perceptron trained to recognize the "10" pattern at 16 Gbps.
	    (a) Input optical signal levels distribution of the 2$^{nd}$ bit of the 2-bit patterns given in the inset.
	    (b) Time traces for the input signal $u(t)$ (blue line) and for the perceptron output $y(t)$ (red dashed line).
	    Dashed black line is the optimal threshold.
	    Gray circles refer to the target response, red circles indicate the best samples at the perceptron output.
	    Vertical stripes indicate the bit slots.
	    (c) Output signal levels distribution of the 2$^{nd}$ bit of the 2-bit patterns given in the inset.
	    The vertical dashed line marks the optimal threshold.
    }
	\label{fig:10levels_trace}
\end{figure}

The first task we consider is the recognition of a pattern of 2 or 3 bits in a pseudo-random bit sequence (PRBS) with a Non Return to Zero (NRZ) modulation format.
This recognition task requires memory.
The perceptron has to output a high state when the target pattern is detected and a low state for all the other cases.
Causality imposes the perceptron to \textit{``wait-and-see"} all the pattern bits before outputting the predictor.
For instance, in the case of a 3-bit pattern, $y$ is delayed by 2 bits and the prediction is aligned with the last bit in the pattern.
The NRZ modulation prevents the device to recognize the symbols with all null bits since the conservation of energy requires the predictor to be identically null, too.\\
A Particle Swarm (PSW) algorithm \cite{part_swarm} has been used to train the perceptron (see method section).
The training has been performed acquiring, for each iteration, a 2 $\mu$s long sequence that corresponds to $3.2\times 10^4$ bits at 16 Gbps, while testing has been done on 10 sequences of the same length.
During both these phases, the algorithm updates the best sampling and threshold values, as it is typical in telecom systems.
In this way, slow drifts due to environmental noise have a smaller impact on the performance of the device during the testing phase.\\
After training, the perceptron performances expressed as the bit error rate (BER) figure at several input signal bit-rates are reported in \Cref{fig:pattern_vs_bitrate}.
Excellent performance is reached for the 2-bit pattern at all bit-rates.
For the case of the 3-bit patterns, the best performance is achieved at either 10 or 16 Gbps, that are the bit-rates closer to the design rate of 20 Gbps (as defined by $\Delta _t$).
The most demanding pattern in terms of memory is "100" because the perceptron has to store the energy of the 1 and release it after two bit slots to output a high level.
We call the rightmost bit of the sequence the reference bit, as it is the bit corresponding to the time the device has to output its predictor.\\
\Cref{fig:10levels_trace} provides a physical insight on the perceptron operation.
The assigned task is to recognize the "10" pattern at 16 Gbps.
\Cref{fig:10levels_trace}(a) reports the distributions of the input levels.
The level of the reference bit is partially affected by the value of the previous one, so that the distributions of "00" and "10" symbols have different mean.
In these cases, the zero-level signal slightly depends on the level of the bit in the past.
This phenomenon is called \textit{intersymbolic interference} and arises from the finite bandwidth of the setup.
On the other hand, the distributions of 1s and 0s of the reference bits are well separated, meaning that the signal information is well conserved.
This is confirmed by the time trace in \Cref{fig:10levels_trace}(b), where the blue continuous line shows the clearly separated high and low levels of the input signal.
In this figure, the signal at the output of the perceptron (predictor) is shown as the red dotted line.
Here, the red dots indicate the best sampling time.
As can be noted the perceptron is able to solve the task for all the instances.
The action of the trained perceptron on the input signal is visible in \Cref{fig:10levels_trace}(c): the distribution of the levels are shifted so that the classes "00", "01", and "11" are well separated from the class "10".
The last is the only one above the threshold (vertical dashed line).
This task highlights the importance of the perceptron memory because it successfully produces a high output even in the presence of a low input state of the reference bit.

\subsection{Delayed XOR task}
We use the n-bit delayed XOR task to investigate the node memory and its non-linear transformation capability \cite{schubert2021local}.
The perceptron has to output the result of the XOR operation between a bit and the n-th previous bit.
\begin{figure}
	\centering
	\includegraphics[scale=1]{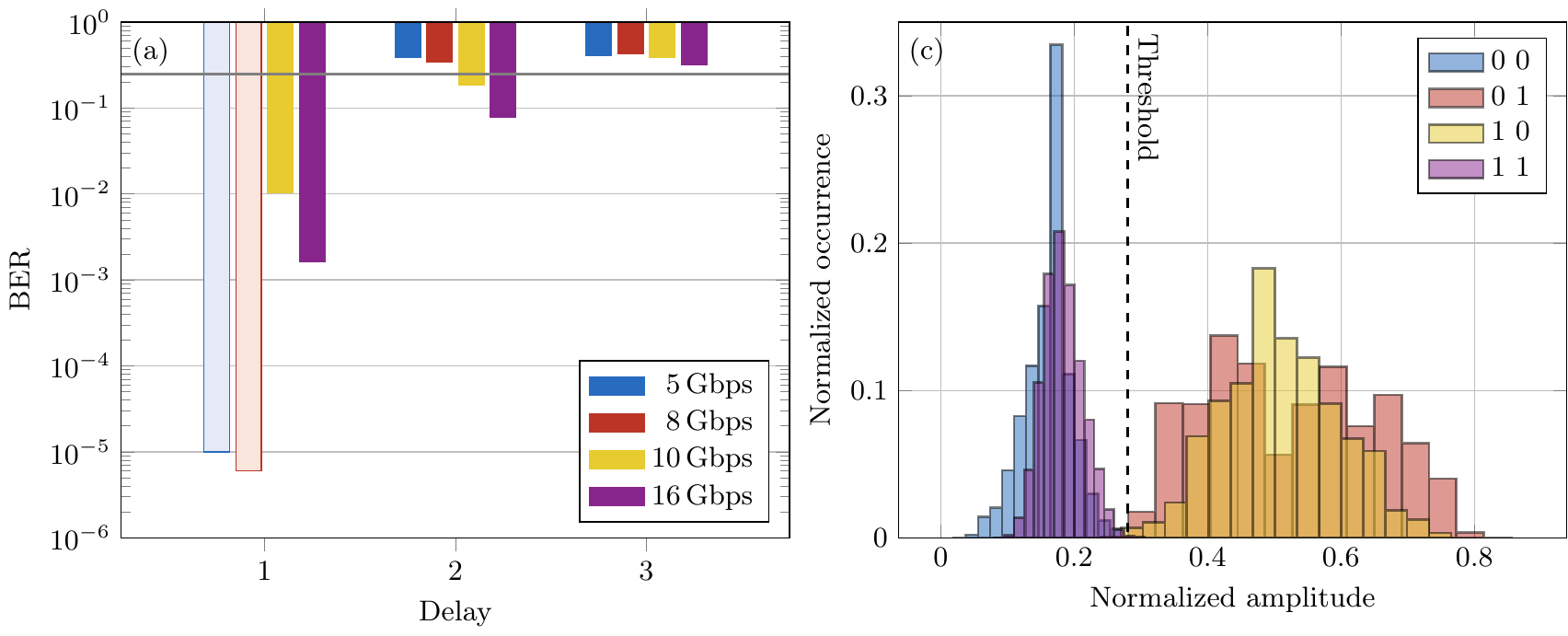}
	\\
	\includegraphics[scale=1]{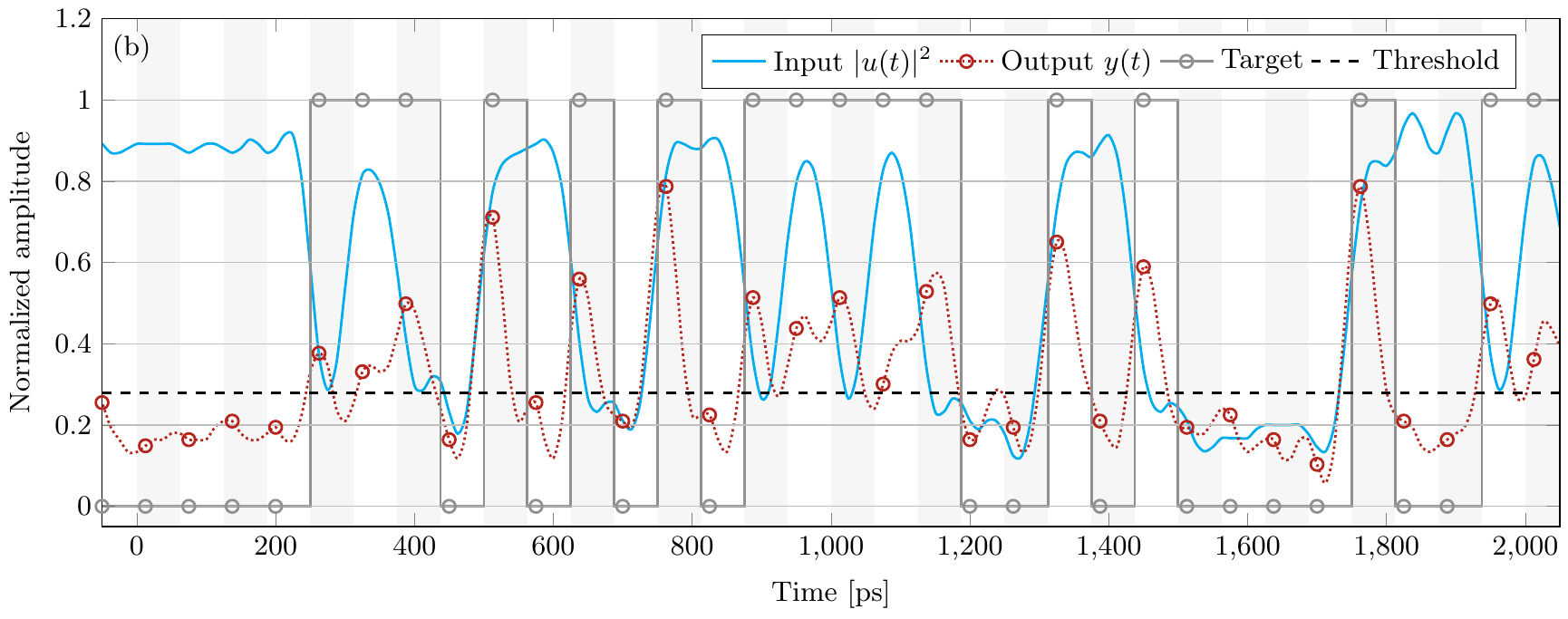}
	\caption{
	    (a) BER for the several bit rates indicated in the inset and XOR bit delay.
	    The BER is indicated with transparent bars when the statistical limit is achieved, i.e., the operation is error-free.
	    The black line refers to the non-linear separability threshold.
	    (b) Input $u$ (blue) and output $y$ (dashed red) time traces for a bit rate of 16 Gbps.
	    The vertical stripes show the bit slots, the horizontal black dashed line indicates the optimal threshold while the red dots the optimal sampling time.
	    Gray dots mark the target.
	    (c) Output signal levels distribution of the 2nd bit of the 2-bit patterns given in the inset.
	    The vertical dashed line marks the optimal threshold.
	    Panels (b) and (c) refer to a bit rate of 16 Gbps.
	    Training and test are carried out at best sampling and threshold values.
    }
	\label{fig:delayed_xor_bitrate_hist_timetrace}
\end{figure}

\Cref{fig:delayed_xor_bitrate_hist_timetrace}(a) shows that, with 1-bit delayed XOR operation, the perceptron has enough memory and non-linearity to perform error-free operation up to 8 Gbps and it achieves a BER $\sim 10^{-3}$ at 16 Gbps.
An example of the output of the perceptron at 16 Gbps is shown in \Cref{fig:delayed_xor_bitrate_hist_timetrace}(b) where the red dots show the best sampling to achieve the desired task with the optimal threshold level indicated by a black dashed horizontal line.
The nonlinear transformation of the input is visible by looking at the output level histograms (\Cref{fig:delayed_xor_bitrate_hist_timetrace}(c)) which show the way the perceptron separates the output levels to perform the desired task (bit-rate 16 Gbps).

\begin{figure}
	\centering
	\includegraphics[scale=1]{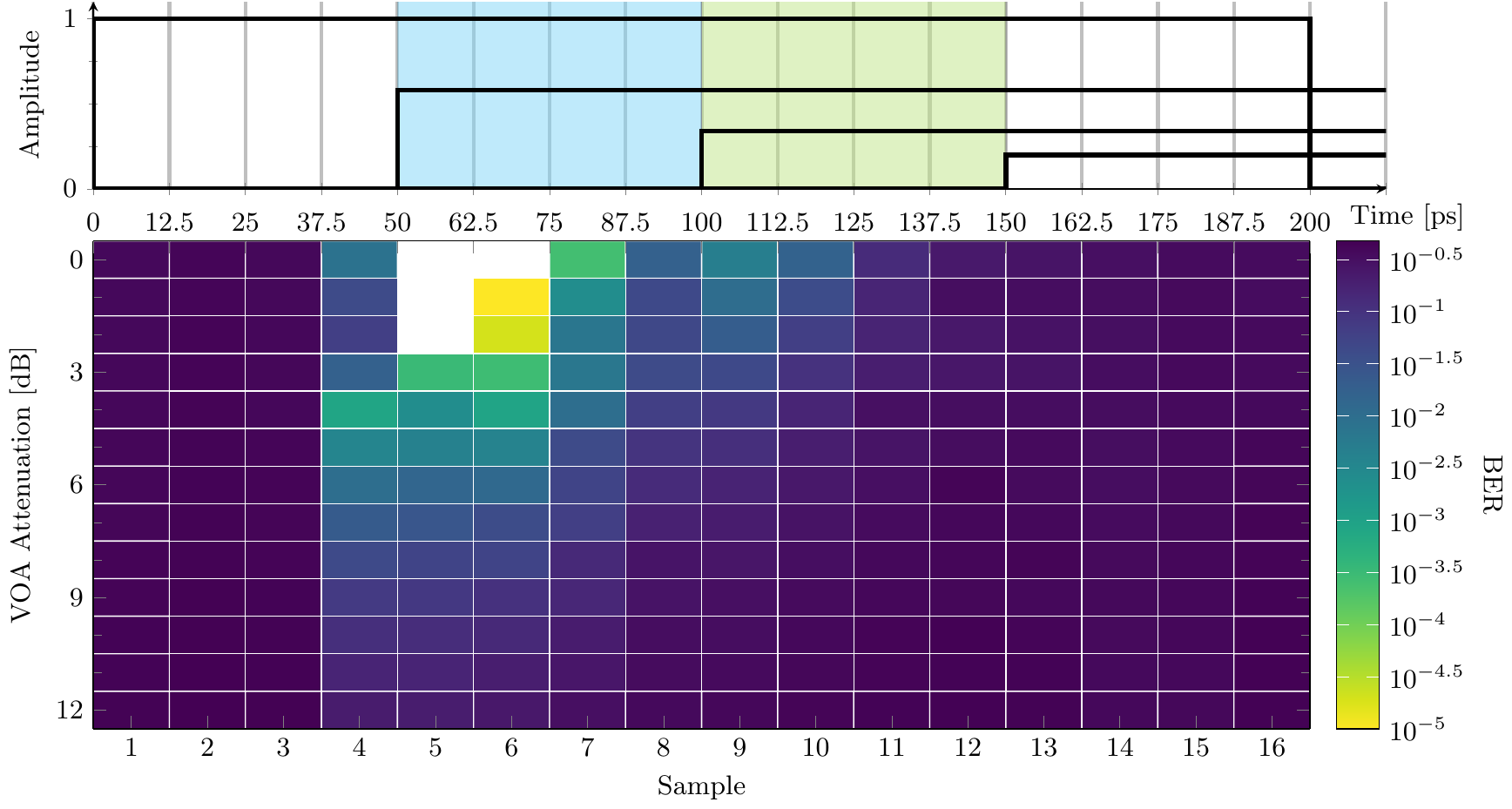}    
	\caption{
	    (Upper) Perceptron delayed inputs $u_i$ (black lines) as a function of time from the begin of the bit for an input equal to $01$.
	    The delayed copies are attenuated accordingly to the spirals losses.
	    Vertical blue and green stripes identify the optimal working region.
	    (bottom) 1-bit delay XOR at 5 Gbps.
	    BER versus the time from the begin of the bit and VOA attenuation.
	    Data at the best BER: VOA attenuation 0 dB, best sampling time 62.5 ps, insertion loss 9 dB, power at the detector 1.1 dBm, input power 4 dBm, BER statistical limit $10^{-5}$.
    }
	\label{fig:delayed_xor_perf_map}
\end{figure}

An important action during the perceptron training phase is the selection of the best sampling time $t_S$, i.e., the best time within the bit slot at which the complex sum (\Cref{Eq:sum}) between the delayed versions $u_k(t)$ of the input $u(t)$ is performed.
In order to clarify the role of the best sampling, we study in details the 1-bit delayed XOR at a bit-rate of 5 Gbps by sampling each bit time slot with $B_{sa}$ = 16 samples (see method section).
In this case, each sample point is separated by 12.5 ps.
Therefore, the perceptron processes four signals delayed by 50 ps at different times of the input bit.
\Cref{fig:delayed_xor_perf_map} reports the BER as a function of the sample number or time computed with respect to the start of the bit, for various input intensity (VOA attenuation).

This measure highlights the system memory by shifting, at steps of 12.5 ps from the start of the bit, the time $t_S$ at which the perceptron outputs the predictor, $y(t_S)$.
The 1-bit delayed XOR task requires a memory long enough to get information on the previous bit, therefore the perceptron might be able to compute until $t_S$ reaches the maximum system memory of 150 ps (the maximum delay between $u_1$ and $u_4$).
At this time, all 4 delayed signals carry information on just the current bit as is shown in the upper panel of \Cref{fig:delayed_xor_perf_map}.
The best BER performance is obtained for $t_S$ = 62.5 ps.
The network trained on the first 3 sampling intervals (i.e., $t_S<$ 50 ps) cannot solve the task.
In this time frame, the sum is \ref{Eq:sum} is performed on $u_1$ coming from the actual bit (note that $u_1$ is the most intense because it is not delayed, i.e., it does not propagate through the spirals) and on $u_2$, $u_3$ and $u_4$ coming from the past bit that are attenuated by the spiral losses.
As shown in \Cref{fig:delayed_example}, the best performance on the XOR task is obtained when $u_1$ and $u_2$ are equal since, by adjusting its relative phase, also their amplitude can be modified in the sum due to the interference between signals coming from the same bit.
This condition is reached at $t_S >$ 50 ps.
The performance of the perceptron degrades when also $u_3$ is populated by the actual bit and fade away when the perceptron memory is overcome.

This simplified scheme is affected by the setup jitter which influences the exact timings shuffling the levels used by the perceptron to process the data.
Furthermore, a symmetric jitter is expected near the bit end but in this case the scenario is even worse since, as the time shift increases, the information coming from longer spirals have more weight to provide the correct predictor but the bits they convey are attenuated and noisy, thus worsening the overall BER (compared to the first half of the bit).

\Cref{fig:delayed_xor_perf_map} shows also the effect of the input signal intensity (represented by the VOA attenuation) on the perceptron performance: the higher the input intensity, the better the BER is.
The input intensity changes the power at the detector, i.e., its signal-to-noise ratio, but not the perceptron transmission regime that remains linear.
This is confirmed by the measured constant insertion loss as the level of the input signal is varied.
It is observed that as the input signal intensity decreases the BER decreases as well.

We performed also the 2-bit delayed and 3-bit delayed XOR tasks (\Cref{fig:delayed_xor_bitrate_hist_timetrace} (a)).
Results show that only for the 2-bit delayed XOR at the highest bit rates (10 and 16 Gbps), the complex perceptron exceed the non-linear separability threshold (solid horizontal line).
This is because only at these rates the two required conditions of equal $u_1$ and $u_2$ with the present bit and, at the same time, of $u_3$ and $u_4$ having the 2-bit delayed bit can be achieved (indeed for these rates the bit periods are 100 and 62.5 ps, respectively).
For the 3-bit delayed XOR task, this condition is never achieved and the perceptron predictions are worse than the non-linear separability threshold.
A rate larger than 20 Gbps would be needed to perform this task.

\subsection{Complex valued optical perceptron modeling}

\begin{figure}
	\centering
	\includegraphics{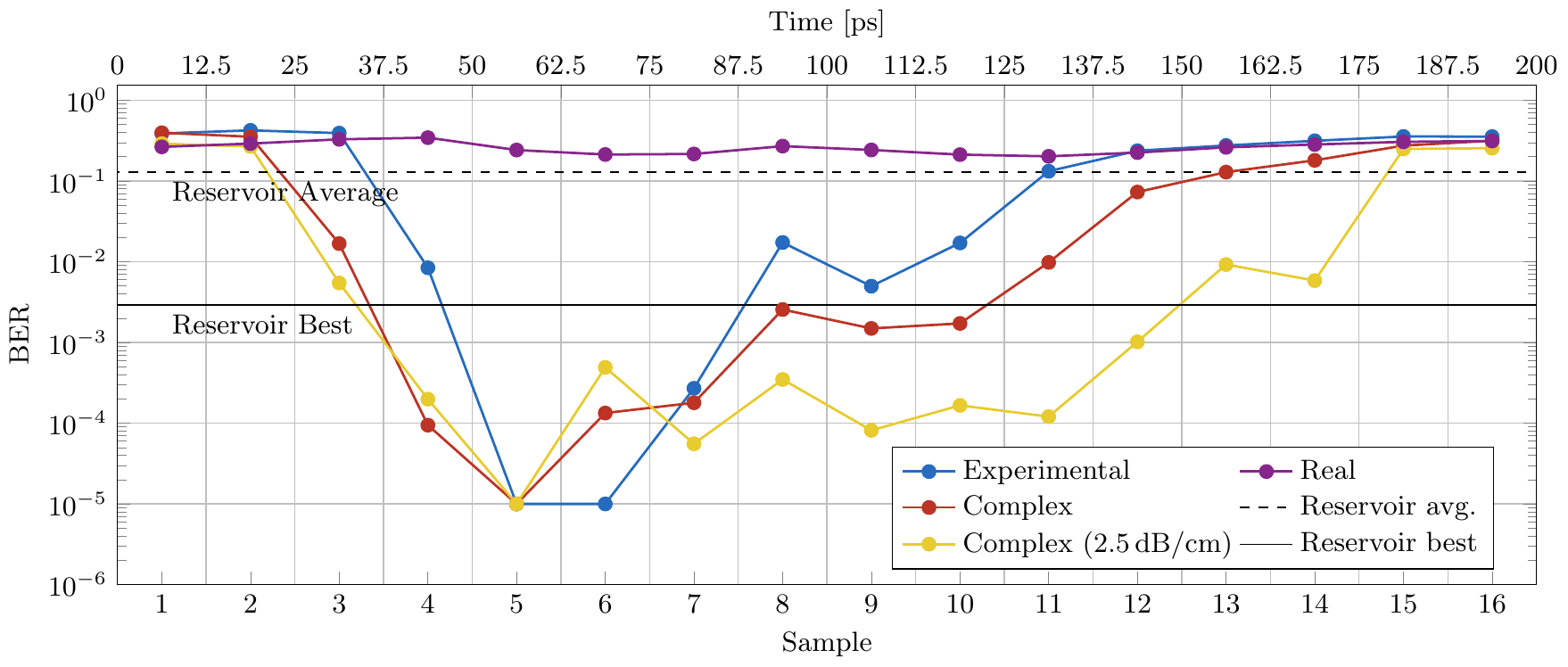}    
	\caption{
	    1-bit delayed XOR.
	    BER as a function of the time from the begin of the bit for the different models indicated in the inset.
	    Simulation parameters: input power 4 dBm, bit-rate 5 Gbps, BER statistical limit $10^{-5}$.
    }
	\label{fig:models}
\end{figure}

To further underline the performances of the complex perceptron, we compare a simulation of the perceptron with the experimental data and we benchmarked it with other schemes based on a similar simple optical circuit.
Three photonic neural networks have been used:
\begin{itemize}
	\item the complex-valued perceptron.
	In this case, the network is modeled as a complex valued perceptron with delayed inputs and $|\cdot|^2$ activation function (\Cref{Eq:sum}).
	The training is performed using the PSW algorithm where only phases are trained.
	This is the model of the measured device.
	\item The real-valued perceptron.
	In this case, the network is modeled as a real valued perceptron.
	Specifically, the modulus square is applied directly on the delayed input copies, i.e., no $|\cdot|^2$ activation function is applied.
	The delayed copies are then weighted with real numbers and summed to produce the prediction.
	The training is performed using a ridge regression and the amplitudes of the weights are changed.
	\item The reservoir computing network with virtual nodes \cite{duport2012all}.
	The complex perceptron is used as a reservoir with random phases.
	The samples of the output are used as virtual nodes.
	The network output is then computed as the weighted sum of the virtual node states, where the optimal weights are found through ridge regression.
	To simulate the random connections of the reservoir, the perceptron output has been determined by giving $4$ random currents to the heaters 10 times.
	The performance is then calculated as the average and best results over the repetitions.
\end{itemize}
The simulated performances of the three networks on the 1-bit delayed XOR task at 5 Gbps and 4 dBm input power are reported in \Cref{fig:models}.
All networks are trained and tested with $B_{sa}$ = 16.
For the reservoir computing network this is the number of virtual nodes.
The experimental data (blue line) is extracted from \Cref{fig:delayed_xor_perf_map} at 2 dB VOA attenuation.

The model of the complex-valued perceptron (red line) reproduces the experimental data (blue line) when a phase noise of $1\%$ is added.
Phase noise accounts for any normally distributed fluctuation of the weights around the trained values.
When the value of propagation losses is lowered to the more reasonable value of $2.5$ dB/cm (yellow line), the performance at longer time is improved.
The real-valued perceptron (violet line) does not solve the 1-bit delayed XOR due to the lack of the non-linearity.
The average performance of the reservoir (dashed line) is not enough to solve the XOR task.
Instead, the best case scenario (solid line) solve the task but at a BER that is order of magnitude worse than the result of complex perceptron.

\subsection*{Phase encoding recognition}
\begin{figure}
	\centering
	\includegraphics[scale=1]{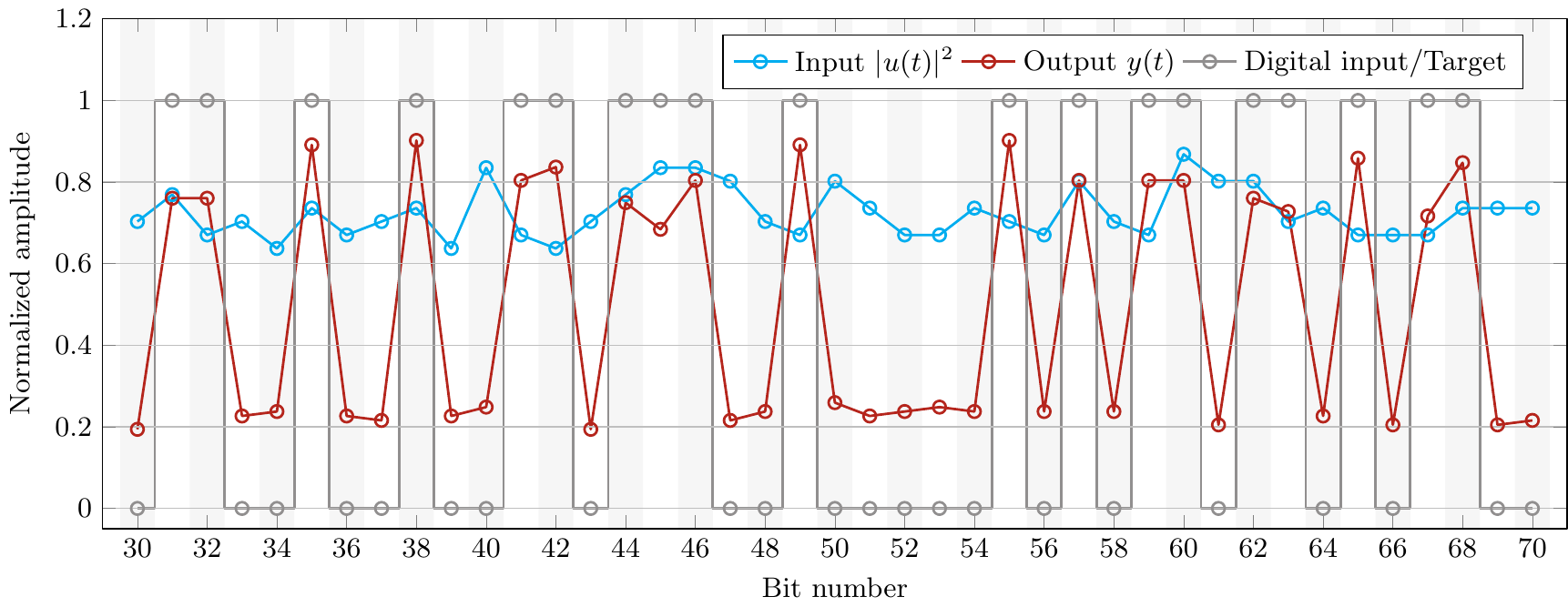}    
	\caption{
	    Input signal intensity (blue), output signal intensity (red) and digital input phase (gray) as a function of the bit number.
    }
	\label{fig:phase_in_to_out}
\end{figure}

A further interesting characteristics of the complex perceptron is the ability to handle pure phase information.
In fact, since the perceptron is based on phase modulation only, it is able to decode phase encoded information, as well.
This is a relevant task which can be used in coherent detection or in protocols of secure communication.
\Cref{fig:phase_in_to_out} reports the results for the trained perceptron which is instructed to translate the phase encoded modulation of the input bit sequence to amplitude modulation of the output sequence.
Black circles show the input bits encoded on the input signal phase (bit 1 phase $\pi$, bit 0 phase 0), red circles show the the signal output.
Decoding is done with zero errors on a 10 Gbps sequence.
The Pearson correlation coefficient between the input intensity and the trained output is 0.002, which proves that the trained output is not determined by the input signal intensity.
Remarkably, the input signal (blue circles) measured on a detector shows only amplitude fluctuations due to noise that are not not correlated with the phase encoded information.

\section{Discussion}

We demonstrate a silicon photonic integrated optical perceptron based on a multiple delayed interferometer that performs logical operation and pattern recognition up to 16 Gbps (limited by our testing system).
The experimental results reflect the complicate interplay between the input bits sequence, the delay lines, and the non-linearly modulated interference effect that outputs the prediction.
For example, the device is expected to perform at its best around a bit duration that is close to the delay of the spiral, i.e., around 16 Gbps.
On the contrary, higher performance is reported at low bit-rates for all the tasks where 1 bit of memory is required(2-bit case).
This fact is related to the training method in which the best sampling time is a free parameter chosen during the training session.
In fact, when the bit duration exceeds the device memory, the best sampling shifts towards the transient between the past and current bits to retain enough memory to solve the tasks and minimize the jitter.
On the 3-bit case, its best performance is obtained at the highest bit rates and depends on the exact input sequence, since a 3-bit pattern uses its full memory.
Once the device is trained, it is a fully all-optical and passive device that does not require any power hungry real time ADC to work.
Furthermore, we demonstrate that it achieves extremely good performance by training only phase weights.
Our scheme enables the training of all system parameters, as in a FFN, thus exploiting the full PIC resources.
It is also able to compare samples from nearby bits to unveil temporal correlations, as in a RC schemes.
Unlike the systems reported in recent works \cite{shen2017deep}, the system computes in the analog domain, the digital-to-analog conversion is carried out only at the perceptron output to read the computed data.
The system memory provided by the spirals is linear thus it can be scaled to higher level without degrading the overall perceptron performance as it happens for nonlinear RC schemes \cite{inubushi2017reservoir}.

The possibility to train the perceptron by applying only phase weights is an important advantage.
In fact the absence of optical attenuation to modulate the signal amplitude, greatly ease its implementation as a node in large networks where propagation losses are relevant.
A proper design of the nodes topology and interconnections will permit to compare samples from bits distant in time and to expand the capability of the network to correlate such information.

\section{Methods}
\subsection*{Experimental setup}
\begin{figure}[t]
    \centering
    \includegraphics[scale=1]{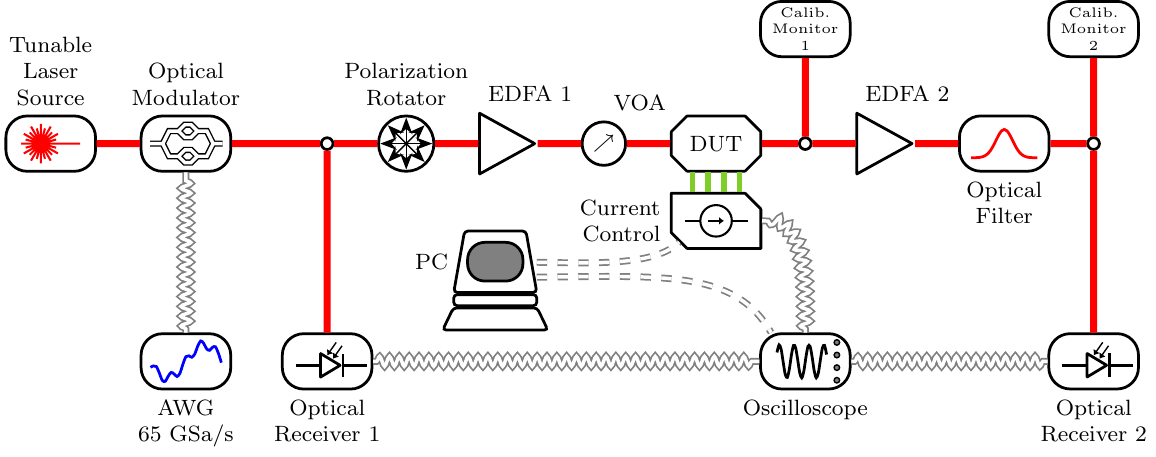}
    \caption{
        Experimental setup.
        The different components are labeled in the figure and explained in the text.
        The complex perceptron is indicated as the DUT (Device Under Test).
        Red lines refer to optical signals, black lines to electrical signals.
    }
    \label{fig:exp_setup}
\end{figure}
The experimental apparatus is sketched in \Cref{fig:exp_setup}.
The laser source is a C-band, CW tunable laser (Pure Photonics) modulated by an electro-optic IQ modulator (IxBlue MXIQ-LN-30) to create the desired input waveform.
The modulator is driven by a 65 GSa Arbitrary Waveform Generator (AWG) from Keysight (KS8195A), whose output is amplified by a high bandwidth amplification stage (IxBlue DR-AN-28-MO), providing the necessary voltage swing of $V_{\pi}$ = 7V to exploit the full dynamic range of the modulator.
A tap of 10{$\%$} is placed at the output of the modulator to monitor the input pump, which is detected by a fast photodiode (RX1, Thorlabs DXM20AF, 20 GHz bandwidth).
The polarization is rotated to match the TE required by the device.
The pump is amplified to a fixed level of 20 dBm by an Erbium Doped Optical Amplifier (EDFA, IPG Photonics) and the input power is regulated by an electronic Variable Optical Attenuation (VOA, VIAVI mVOA-C1).
As a consequence, this VOA also regulates the power at output photodiode (RX2, Thorlabs DXM20AF) that ranges between 4-10 dBm.
The power level at the device output is monitored through a tap of $1\%$ and a low noise photodiode (M1, Viavi mOPM).
Light is amplified by a second EDFA (Thorlabs EDFA100s) operating at constant current.
A tunable optical band-pass filter (25 GHz bandwidth) removes the broadband ASE noise prior to reach RX2.
Another photodiode (M2, mOPM) monitors precisely the power reaching RX2.
A $2 \times 80$ GSa/s oscilloscope with 16 GHz  analog bandwidth (LeCroy SDA 816Zi-A with interleavers) records the input  $u_{in}(t)$ and the output $u_{out}(t)$ waveforms.
A computer controls the current flowing in the 4 heaters integrated onto the device through an 8 channels current generator (Qontrol Q8iv).
An extra current output is sent to the trigger input port of the oscilloscope as time reference.
The training algorithm runs on the PC that modulates the currents accordingly to the acquired input and output waveforms.

\subsection*{Integrated device}
The device has been fabricated on a CMOS facility on silicon-on-insulator wafer with a device layer thickness of 220 nm (iSiPP50G technology process by IMEC thorugh a MPW scheme).
Silicon waveguides (WG) are embedded in a silica cladding, and have a width of 450 nm to ensure single mode operation on both polarizations.
TiN tracks deposited on top of the silica cladding enables local thermal tuning of WG effective index.
A CMOS packaging holds and thermalize at 21 {$\,^{\circ}\mathrm{C}$} the photonic chip using a PID controller and a Peltier heater.
The measured single mode WG propagation loss measured through spirals of different lengths is $\sim$ 6 dB/cm.
This value is highly above the 2 dB/cm average performance of IMEC iSiPP50G process.
Therefore, we consider the device performances to be sub-optimal.
Fiber-to-chip coupling is ensured by grating couplers whose measured insertion loss is $\sim$ 3.8 dB/grating at the maximum transmission wavelength of 1560 nm.\\

\subsection*{Perceptron training and testing procedures}
We used a Particle Swarm (PSW) algorithm to train the optical node \cite{part_swarm}.
Despite its stochastic nature, the limited number of parameters of our system makes it the easiest way to optimize our perceptron as it does not require direct access to the fields in the different waveguides and the activation function can be considered without any approximation.
More complex networks would benefit of gradient-descent, back-propagation-like algorithms, which assure faster convergence and have been reported for photonic implementation of FFN in \cite{hughes2018training}.

The training is carried out by a PC that elaborates the output signal and regulates the current control following the PSW algorithm, as shown in \Cref{fig:exp_setup}.
The typical training time is few tens of seconds and is mainly limited by oscilloscope-PC data exchange.
An FPGA/ASIC implementation of the PSW would greatly decrease this time, which then would become limited only by the speed of the mechanism used to change the weights (MHz in our system).
The input signal is a 8-bit Pseudo Random Binary Sequence (PRBS) that is amplitude modulated NRZ (not return to zero) between 5 and 16 Gbps, with an extinction ratio and a SNR of 7 dB and 14 dB, respectively.
The delay between the input (RX1) and output trace (RX2) is found by cross correlating the traces.
This operation is performed every time the bit rate of the experiment is changed.
The operations of the training phase, performed at each PSW iteration, are:
\begin{itemize}
    \item current values are generated by the PSW and are applied by the current controller to the heaters while the trigger signal is sent to the oscilloscope.
    At this time, only 3 phases are trained to avoid redundancy in the loss function due to the system intrinsic periodicity of $2\pi$
    \item to take into account the dynamics of the heater controllers, we use a delay of 1 ms after the trigger signal before the oscilloscope acquires the input ($x^{RX1}_j$) and output ($x^{RX2}_j$) traces sampled at 80 GSa/s for a total of M = 160 kSa, j = 1,..., M.
    The number of processed bits depends on the bit rate, for instance, at 16 Gbps there are a total of $10^4$ bit and 5 Samples/bit.
    \item the PC aligns the traces using the already measured delay for that specific bit rate.
    \item $x^{RX1}_j$ is digitized ($X^{RX1}_l$) using its average as threshold and selecting the central bit sample.
    The bit index $l$ goes from $1$ to $M/B_{Sa}$, where $B_{Sa} = \frac{80\times 10^9}{\text{bit rate}}$ is the number of samples per bit.
    The target binary sequence $T_l$ is then calculated by applying the task function to $X_l^{RX1}$.
    \item $x^{RX2}_j$ is digitized ($X^{RX2}_{l}$) using a variable threshold $r = {w_{min}^{RX2}, ..., w_{max}^{RX2}}$ spanning the signal dynamic range of the samples $n = 1, ..., B_{sa}$.
    \item the loss function is calculated as  $L_{rn} = \sum_{l=1}^{M/B_{sa}}{ |X^{RX2}_l - T_l|^2 } $.
    The minimum of $L_{rn}$ is considered by selecting the best threshold $r_b$ and the best sampling position $n$.
    The BER is calculated as $BER = \frac{L_{rn}}{M B_{Sa}}$.
\end{itemize}
The algorithm iterates these steps either until error free or the max iteration number are reached.
To note that each input trace provided to the perceptron during the training is unique since the various traces are all corrupted by the experimental noise (such as the detector thermal and shot noise, the phase noise on the weights due to the micro-thermal variation on the spirals, and the time jitters of the AWG/oscilloscope sampling times).
Such noise slows down the training process but helps in avoiding over-fitting.
As a result of the train the 3 optimal currents found are used in the  testing phase where 10 new traces are acquired and processed.
The test BER is the average of the BER calculated for each acquisition.

\section*{Acknowledgements}
This project has received funding from the European Research Council (ERC) under the European Union's Horizon 2020 research and innovation program (grant agreement No 788793, BACKUP and grant agreement No 963463, ALPI), and from the MIUR under the project PRIN PELM (20177 PSCKT).

\section*{Author Contributions} 
M.M. conceived the experiment.
M.M. and D.B. designed the device, made the experiments and analyzed the data.
M.M. simulated the experiment.
P.B. and L.P. supervised the project.
All authors contributed to the discussion of the results and to the writing of the manuscript.

\section*{Competing Interests statement}
M.M., P.B. and L.P. declare that they have filed a patent on the device described in this paper.

\bibliography{main}

\begin{thebibliography}{20}%
\makeatletter
\providecommand \@ifxundefined [1]{%
 \@ifx{#1\undefined}
}%
\providecommand \@ifnum [1]{%
 \ifnum #1\expandafter \@firstoftwo
 \else \expandafter \@secondoftwo
 \fi
}%
\providecommand \@ifx [1]{%
 \ifx #1\expandafter \@firstoftwo
 \else \expandafter \@secondoftwo
 \fi
}%
\providecommand \natexlab [1]{#1}%
\providecommand \enquote  [1]{``#1''}%
\providecommand \bibnamefont  [1]{#1}%
\providecommand \bibfnamefont [1]{#1}%
\providecommand \citenamefont [1]{#1}%
\providecommand \href@noop [0]{\@secondoftwo}%
\providecommand \href [0]{\begingroup \@sanitize@url \@href}%
\providecommand \@href[1]{\@@startlink{#1}\@@href}%
\providecommand \@@href[1]{\endgroup#1\@@endlink}%
\providecommand \@sanitize@url [0]{\catcode `\\12\catcode `\$12\catcode
  `\&12\catcode `\#12\catcode `\^12\catcode `\_12\catcode `\%12\relax}%
\providecommand \@@startlink[1]{}%
\providecommand \@@endlink[0]{}%
\providecommand \url  [0]{\begingroup\@sanitize@url \@url }%
\providecommand \@url [1]{\endgroup\@href {#1}{\urlprefix }}%
\providecommand \urlprefix  [0]{URL }%
\providecommand \Eprint [0]{\href }%
\providecommand \doibase [0]{http://dx.doi.org/}%
\providecommand \selectlanguage [0]{\@gobble}%
\providecommand \bibinfo  [0]{\@secondoftwo}%
\providecommand \bibfield  [0]{\@secondoftwo}%
\providecommand \translation [1]{[#1]}%
\providecommand \BibitemOpen [0]{}%
\providecommand \bibitemStop [0]{}%
\providecommand \bibitemNoStop [0]{.\EOS\space}%
\providecommand \EOS [0]{\spacefactor3000\relax}%
\providecommand \BibitemShut  [1]{\csname bibitem#1\endcsname}%
\let\auto@bib@innerbib\@empty
\bibitem [{\citenamefont {Xu}\ \emph {et~al.}(2021)\citenamefont {Xu},
  \citenamefont {Tan}, \citenamefont {Corcoran}, \citenamefont {Wu},
  \citenamefont {Boes}, \citenamefont {Nguyen}, \citenamefont {Chu},
  \citenamefont {Little}, \citenamefont {Hicks}, \citenamefont {Morandotti}
  \emph {et~al.}}]{xu202111}%
  \BibitemOpen
  \bibfield  {author} {\bibinfo {author} {\bibfnamefont {X.}~\bibnamefont
  {Xu}}, \bibinfo {author} {\bibfnamefont {M.}~\bibnamefont {Tan}}, \bibinfo
  {author} {\bibfnamefont {B.}~\bibnamefont {Corcoran}}, \bibinfo {author}
  {\bibfnamefont {J.}~\bibnamefont {Wu}}, \bibinfo {author} {\bibfnamefont
  {A.}~\bibnamefont {Boes}}, \bibinfo {author} {\bibfnamefont {T.~G.}\
  \bibnamefont {Nguyen}}, \bibinfo {author} {\bibfnamefont {S.~T.}\
  \bibnamefont {Chu}}, \bibinfo {author} {\bibfnamefont {B.~E.}\ \bibnamefont
  {Little}}, \bibinfo {author} {\bibfnamefont {D.~G.}\ \bibnamefont {Hicks}},
  \bibinfo {author} {\bibfnamefont {R.}~\bibnamefont {Morandotti}},  \emph
  {et~al.},\ }\href@noop {} {\bibfield  {journal} {\bibinfo  {journal}
  {Nature}\ }\textbf {\bibinfo {volume} {589}},\ \bibinfo {pages} {44}
  (\bibinfo {year} {2021})}\BibitemShut {NoStop}%
\bibitem [{\citenamefont {Feldmann}\ \emph {et~al.}(2021)\citenamefont
  {Feldmann}, \citenamefont {Youngblood}, \citenamefont {Karpov}, \citenamefont
  {Gehring}, \citenamefont {Li}, \citenamefont {Stappers}, \citenamefont
  {Le~Gallo}, \citenamefont {Fu}, \citenamefont {Lukashchuk}, \citenamefont
  {Raja} \emph {et~al.}}]{feldmann2021parallel}%
  \BibitemOpen
  \bibfield  {author} {\bibinfo {author} {\bibfnamefont {J.}~\bibnamefont
  {Feldmann}}, \bibinfo {author} {\bibfnamefont {N.}~\bibnamefont
  {Youngblood}}, \bibinfo {author} {\bibfnamefont {M.}~\bibnamefont {Karpov}},
  \bibinfo {author} {\bibfnamefont {H.}~\bibnamefont {Gehring}}, \bibinfo
  {author} {\bibfnamefont {X.}~\bibnamefont {Li}}, \bibinfo {author}
  {\bibfnamefont {M.}~\bibnamefont {Stappers}}, \bibinfo {author}
  {\bibfnamefont {M.}~\bibnamefont {Le~Gallo}}, \bibinfo {author}
  {\bibfnamefont {X.}~\bibnamefont {Fu}}, \bibinfo {author} {\bibfnamefont
  {A.}~\bibnamefont {Lukashchuk}}, \bibinfo {author} {\bibfnamefont {A.~S.}\
  \bibnamefont {Raja}},  \emph {et~al.},\ }\href@noop {} {\bibfield  {journal}
  {\bibinfo  {journal} {Nature}\ }\textbf {\bibinfo {volume} {589}},\ \bibinfo
  {pages} {52} (\bibinfo {year} {2021})}\BibitemShut {NoStop}%
\bibitem [{\citenamefont {Vandoorne}\ \emph {et~al.}(2011)\citenamefont
  {Vandoorne}, \citenamefont {Dambre}, \citenamefont {Verstraeten},
  \citenamefont {Schrauwen},\ and\ \citenamefont
  {Bienstman}}]{vandoorne2011parallel}%
  \BibitemOpen
  \bibfield  {author} {\bibinfo {author} {\bibfnamefont {K.}~\bibnamefont
  {Vandoorne}}, \bibinfo {author} {\bibfnamefont {J.}~\bibnamefont {Dambre}},
  \bibinfo {author} {\bibfnamefont {D.}~\bibnamefont {Verstraeten}}, \bibinfo
  {author} {\bibfnamefont {B.}~\bibnamefont {Schrauwen}}, \ and\ \bibinfo
  {author} {\bibfnamefont {P.}~\bibnamefont {Bienstman}},\ }\href@noop {}
  {\bibfield  {journal} {\bibinfo  {journal} {IEEE transactions on neural
  networks}\ }\textbf {\bibinfo {volume} {22}},\ \bibinfo {pages} {1469}
  (\bibinfo {year} {2011})}\BibitemShut {NoStop}%
\bibitem [{\citenamefont {Katumba}\ \emph {et~al.}(2019)\citenamefont
  {Katumba}, \citenamefont {Yin}, \citenamefont {Dambre},\ and\ \citenamefont
  {Bienstman}}]{katumba2019neuromorphic}%
  \BibitemOpen
  \bibfield  {author} {\bibinfo {author} {\bibfnamefont {A.}~\bibnamefont
  {Katumba}}, \bibinfo {author} {\bibfnamefont {X.}~\bibnamefont {Yin}},
  \bibinfo {author} {\bibfnamefont {J.}~\bibnamefont {Dambre}}, \ and\ \bibinfo
  {author} {\bibfnamefont {P.}~\bibnamefont {Bienstman}},\ }\href@noop {}
  {\bibfield  {journal} {\bibinfo  {journal} {Journal of Lightwave Technology}\
  }\textbf {\bibinfo {volume} {37}},\ \bibinfo {pages} {2232} (\bibinfo {year}
  {2019})}\BibitemShut {NoStop}%
\bibitem [{\citenamefont {Argyris}\ \emph {et~al.}(2018)\citenamefont
  {Argyris}, \citenamefont {Bueno},\ and\ \citenamefont
  {Fischer}}]{argyris2018photonic}%
  \BibitemOpen
  \bibfield  {author} {\bibinfo {author} {\bibfnamefont {A.}~\bibnamefont
  {Argyris}}, \bibinfo {author} {\bibfnamefont {J.}~\bibnamefont {Bueno}}, \
  and\ \bibinfo {author} {\bibfnamefont {I.}~\bibnamefont {Fischer}},\
  }\href@noop {} {\bibfield  {journal} {\bibinfo  {journal} {Scientific
  reports}\ }\textbf {\bibinfo {volume} {8}},\ \bibinfo {pages} {1} (\bibinfo
  {year} {2018})}\BibitemShut {NoStop}%
\bibitem [{\citenamefont {Harris}\ \emph {et~al.}(2018)\citenamefont {Harris},
  \citenamefont {Carolan}, \citenamefont {Bunandar}, \citenamefont {Prabhu},
  \citenamefont {Hochberg}, \citenamefont {Baehr-Jones}, \citenamefont {Fanto},
  \citenamefont {Smith}, \citenamefont {Tison}, \citenamefont {Alsing} \emph
  {et~al.}}]{harris2018linear}%
  \BibitemOpen
  \bibfield  {author} {\bibinfo {author} {\bibfnamefont {N.~C.}\ \bibnamefont
  {Harris}}, \bibinfo {author} {\bibfnamefont {J.}~\bibnamefont {Carolan}},
  \bibinfo {author} {\bibfnamefont {D.}~\bibnamefont {Bunandar}}, \bibinfo
  {author} {\bibfnamefont {M.}~\bibnamefont {Prabhu}}, \bibinfo {author}
  {\bibfnamefont {M.}~\bibnamefont {Hochberg}}, \bibinfo {author}
  {\bibfnamefont {T.}~\bibnamefont {Baehr-Jones}}, \bibinfo {author}
  {\bibfnamefont {M.~L.}\ \bibnamefont {Fanto}}, \bibinfo {author}
  {\bibfnamefont {A.~M.}\ \bibnamefont {Smith}}, \bibinfo {author}
  {\bibfnamefont {C.~C.}\ \bibnamefont {Tison}}, \bibinfo {author}
  {\bibfnamefont {P.~M.}\ \bibnamefont {Alsing}},  \emph {et~al.},\ }\href@noop
  {} {\bibfield  {journal} {\bibinfo  {journal} {Optica}\ }\textbf {\bibinfo
  {volume} {5}},\ \bibinfo {pages} {1623} (\bibinfo {year} {2018})}\BibitemShut
  {NoStop}%
\bibitem [{\citenamefont {Peng}\ \emph {et~al.}(2018)\citenamefont {Peng},
  \citenamefont {Nahmias}, \citenamefont {De~Lima}, \citenamefont {Tait},\ and\
  \citenamefont {Shastri}}]{peng2018neuromorphic}%
  \BibitemOpen
  \bibfield  {author} {\bibinfo {author} {\bibfnamefont {H.-T.}\ \bibnamefont
  {Peng}}, \bibinfo {author} {\bibfnamefont {M.~A.}\ \bibnamefont {Nahmias}},
  \bibinfo {author} {\bibfnamefont {T.~F.}\ \bibnamefont {De~Lima}}, \bibinfo
  {author} {\bibfnamefont {A.~N.}\ \bibnamefont {Tait}}, \ and\ \bibinfo
  {author} {\bibfnamefont {B.~J.}\ \bibnamefont {Shastri}},\ }\href@noop {}
  {\bibfield  {journal} {\bibinfo  {journal} {IEEE Journal of Selected Topics
  in Quantum Electronics}\ }\textbf {\bibinfo {volume} {24}},\ \bibinfo {pages}
  {1} (\bibinfo {year} {2018})}\BibitemShut {NoStop}%
\bibitem [{\citenamefont {Genty}\ \emph {et~al.}(2020)\citenamefont {Genty},
  \citenamefont {Salmela}, \citenamefont {Dudley}, \citenamefont {Brunner},
  \citenamefont {Kokhanovskiy}, \citenamefont {Kobtsev},\ and\ \citenamefont
  {Turitsyn}}]{genty2020machine}%
  \BibitemOpen
  \bibfield  {author} {\bibinfo {author} {\bibfnamefont {G.}~\bibnamefont
  {Genty}}, \bibinfo {author} {\bibfnamefont {L.}~\bibnamefont {Salmela}},
  \bibinfo {author} {\bibfnamefont {J.~M.}\ \bibnamefont {Dudley}}, \bibinfo
  {author} {\bibfnamefont {D.}~\bibnamefont {Brunner}}, \bibinfo {author}
  {\bibfnamefont {A.}~\bibnamefont {Kokhanovskiy}}, \bibinfo {author}
  {\bibfnamefont {S.}~\bibnamefont {Kobtsev}}, \ and\ \bibinfo {author}
  {\bibfnamefont {S.~K.}\ \bibnamefont {Turitsyn}},\ }\href@noop {} {\bibfield
  {journal} {\bibinfo  {journal} {Nature Photonics}\ ,\ \bibinfo {pages} {1}}
  (\bibinfo {year} {2020})}\BibitemShut {NoStop}%
\bibitem [{\citenamefont {Hughes}\ \emph {et~al.}(2018)\citenamefont {Hughes},
  \citenamefont {Minkov}, \citenamefont {Shi},\ and\ \citenamefont
  {Fan}}]{hughes2018training}%
  \BibitemOpen
  \bibfield  {author} {\bibinfo {author} {\bibfnamefont {T.~W.}\ \bibnamefont
  {Hughes}}, \bibinfo {author} {\bibfnamefont {M.}~\bibnamefont {Minkov}},
  \bibinfo {author} {\bibfnamefont {Y.}~\bibnamefont {Shi}}, \ and\ \bibinfo
  {author} {\bibfnamefont {S.}~\bibnamefont {Fan}},\ }\href@noop {} {\bibfield
  {journal} {\bibinfo  {journal} {Optica}\ }\textbf {\bibinfo {volume} {5}},\
  \bibinfo {pages} {864} (\bibinfo {year} {2018})}\BibitemShut {NoStop}%
\bibitem [{\citenamefont {Xu}\ \emph {et~al.}(2020)\citenamefont {Xu},
  \citenamefont {Tan}, \citenamefont {Corcoran}, \citenamefont {Wu},
  \citenamefont {Nguyen}, \citenamefont {Boes}, \citenamefont {Chu},
  \citenamefont {Little}, \citenamefont {Morandotti}, \citenamefont {Mitchell}
  \emph {et~al.}}]{xu2020photonic}%
  \BibitemOpen
  \bibfield  {author} {\bibinfo {author} {\bibfnamefont {X.}~\bibnamefont
  {Xu}}, \bibinfo {author} {\bibfnamefont {M.}~\bibnamefont {Tan}}, \bibinfo
  {author} {\bibfnamefont {B.}~\bibnamefont {Corcoran}}, \bibinfo {author}
  {\bibfnamefont {J.}~\bibnamefont {Wu}}, \bibinfo {author} {\bibfnamefont
  {T.~G.}\ \bibnamefont {Nguyen}}, \bibinfo {author} {\bibfnamefont
  {A.}~\bibnamefont {Boes}}, \bibinfo {author} {\bibfnamefont {S.~T.}\
  \bibnamefont {Chu}}, \bibinfo {author} {\bibfnamefont {B.~E.}\ \bibnamefont
  {Little}}, \bibinfo {author} {\bibfnamefont {R.}~\bibnamefont {Morandotti}},
  \bibinfo {author} {\bibfnamefont {A.}~\bibnamefont {Mitchell}},  \emph
  {et~al.},\ }\href@noop {} {\bibfield  {journal} {\bibinfo  {journal} {Laser
  \& Photonics Reviews}\ }\textbf {\bibinfo {volume} {14}},\ \bibinfo {pages}
  {2000070} (\bibinfo {year} {2020})}\BibitemShut {NoStop}%
\bibitem [{\citenamefont {Carroll}(2020)}]{carroll2020reservoir}%
  \BibitemOpen
  \bibfield  {author} {\bibinfo {author} {\bibfnamefont {T.~L.}\ \bibnamefont
  {Carroll}},\ }\href@noop {} {\bibfield  {journal} {\bibinfo  {journal}
  {Chaos: An Interdisciplinary Journal of Nonlinear Science}\ }\textbf
  {\bibinfo {volume} {30}},\ \bibinfo {pages} {121109} (\bibinfo {year}
  {2020})}\BibitemShut {NoStop}%
\bibitem [{\citenamefont {Appeltant}\ \emph {et~al.}(2011)\citenamefont
  {Appeltant}, \citenamefont {Soriano}, \citenamefont {Van~der Sande},
  \citenamefont {Danckaert}, \citenamefont {Massar}, \citenamefont {Dambre},
  \citenamefont {Schrauwen}, \citenamefont {Mirasso},\ and\ \citenamefont
  {Fischer}}]{appeltant2011information}%
  \BibitemOpen
  \bibfield  {author} {\bibinfo {author} {\bibfnamefont {L.}~\bibnamefont
  {Appeltant}}, \bibinfo {author} {\bibfnamefont {M.~C.}\ \bibnamefont
  {Soriano}}, \bibinfo {author} {\bibfnamefont {G.}~\bibnamefont {Van~der
  Sande}}, \bibinfo {author} {\bibfnamefont {J.}~\bibnamefont {Danckaert}},
  \bibinfo {author} {\bibfnamefont {S.}~\bibnamefont {Massar}}, \bibinfo
  {author} {\bibfnamefont {J.}~\bibnamefont {Dambre}}, \bibinfo {author}
  {\bibfnamefont {B.}~\bibnamefont {Schrauwen}}, \bibinfo {author}
  {\bibfnamefont {C.~R.}\ \bibnamefont {Mirasso}}, \ and\ \bibinfo {author}
  {\bibfnamefont {I.}~\bibnamefont {Fischer}},\ }\href@noop {} {\bibfield
  {journal} {\bibinfo  {journal} {Nature communications}\ }\textbf {\bibinfo
  {volume} {2}},\ \bibinfo {pages} {1} (\bibinfo {year} {2011})}\BibitemShut
  {NoStop}%
\bibitem [{\citenamefont {Duport}\ \emph {et~al.}(2012)\citenamefont {Duport},
  \citenamefont {Schneider}, \citenamefont {Smerieri}, \citenamefont
  {Haelterman},\ and\ \citenamefont {Massar}}]{duport2012all}%
  \BibitemOpen
  \bibfield  {author} {\bibinfo {author} {\bibfnamefont {F.}~\bibnamefont
  {Duport}}, \bibinfo {author} {\bibfnamefont {B.}~\bibnamefont {Schneider}},
  \bibinfo {author} {\bibfnamefont {A.}~\bibnamefont {Smerieri}}, \bibinfo
  {author} {\bibfnamefont {M.}~\bibnamefont {Haelterman}}, \ and\ \bibinfo
  {author} {\bibfnamefont {S.}~\bibnamefont {Massar}},\ }\href@noop {}
  {\bibfield  {journal} {\bibinfo  {journal} {Optics express}\ }\textbf
  {\bibinfo {volume} {20}},\ \bibinfo {pages} {22783} (\bibinfo {year}
  {2012})}\BibitemShut {NoStop}%
\bibitem [{\citenamefont {Vandoorne}\ \emph {et~al.}(2014)\citenamefont
  {Vandoorne}, \citenamefont {Mechet}, \citenamefont {Van~Vaerenbergh},
  \citenamefont {Fiers}, \citenamefont {Morthier}, \citenamefont {Verstraeten},
  \citenamefont {Schrauwen}, \citenamefont {Dambre},\ and\ \citenamefont
  {Bienstman}}]{vandoorne2014experimental}%
  \BibitemOpen
  \bibfield  {author} {\bibinfo {author} {\bibfnamefont {K.}~\bibnamefont
  {Vandoorne}}, \bibinfo {author} {\bibfnamefont {P.}~\bibnamefont {Mechet}},
  \bibinfo {author} {\bibfnamefont {T.}~\bibnamefont {Van~Vaerenbergh}},
  \bibinfo {author} {\bibfnamefont {M.}~\bibnamefont {Fiers}}, \bibinfo
  {author} {\bibfnamefont {G.}~\bibnamefont {Morthier}}, \bibinfo {author}
  {\bibfnamefont {D.}~\bibnamefont {Verstraeten}}, \bibinfo {author}
  {\bibfnamefont {B.}~\bibnamefont {Schrauwen}}, \bibinfo {author}
  {\bibfnamefont {J.}~\bibnamefont {Dambre}}, \ and\ \bibinfo {author}
  {\bibfnamefont {P.}~\bibnamefont {Bienstman}},\ }\href@noop {} {\bibfield
  {journal} {\bibinfo  {journal} {Nature communications}\ }\textbf {\bibinfo
  {volume} {5}},\ \bibinfo {pages} {1} (\bibinfo {year} {2014})}\BibitemShut
  {NoStop}%
\bibitem [{\citenamefont {Tait}\ \emph {et~al.}(2014)\citenamefont {Tait},
  \citenamefont {Nahmias}, \citenamefont {Shastri},\ and\ \citenamefont
  {Prucnal}}]{tait2014broadcast}%
  \BibitemOpen
  \bibfield  {author} {\bibinfo {author} {\bibfnamefont {A.~N.}\ \bibnamefont
  {Tait}}, \bibinfo {author} {\bibfnamefont {M.~A.}\ \bibnamefont {Nahmias}},
  \bibinfo {author} {\bibfnamefont {B.~J.}\ \bibnamefont {Shastri}}, \ and\
  \bibinfo {author} {\bibfnamefont {P.~R.}\ \bibnamefont {Prucnal}},\
  }\href@noop {} {\bibfield  {journal} {\bibinfo  {journal} {Journal of
  Lightwave Technology}\ }\textbf {\bibinfo {volume} {32}},\ \bibinfo {pages}
  {4029} (\bibinfo {year} {2014})}\BibitemShut {NoStop}%
\bibitem [{\citenamefont {Rosenblatt}(1957)}]{rosenblatt1957perceptron}%
  \BibitemOpen
  \bibfield  {author} {\bibinfo {author} {\bibfnamefont {F.}~\bibnamefont
  {Rosenblatt}},\ }\href@noop {} {\emph {\bibinfo {title} {The perceptron, a
  perceiving and recognizing automaton Project Para}}}\ (\bibinfo  {publisher}
  {Cornell Aeronautical Laboratory},\ \bibinfo {year} {1957})\BibitemShut
  {NoStop}%
\bibitem [{\citenamefont {Bansal}\ \emph {et~al.}(2019)\citenamefont {Bansal},
  \citenamefont {Singh},\ and\ \citenamefont {Pal}}]{part_swarm}%
  \BibitemOpen
  \bibfield  {author} {\bibinfo {author} {\bibfnamefont {J.~C.}\ \bibnamefont
  {Bansal}}, \bibinfo {author} {\bibfnamefont {P.~K.}\ \bibnamefont {Singh}}, \
  and\ \bibinfo {author} {\bibfnamefont {N.~R.}\ \bibnamefont {Pal}},\
  }\href@noop {} {\emph {\bibinfo {title} {Evolutionary and swarm intelligence
  algorithms}}}\ (\bibinfo  {publisher} {Springer},\ \bibinfo {year}
  {2019})\BibitemShut {NoStop}%
\bibitem [{\citenamefont {Schubert}\ and\ \citenamefont
  {Gros}(2021)}]{schubert2021local}%
  \BibitemOpen
  \bibfield  {author} {\bibinfo {author} {\bibfnamefont {F.}~\bibnamefont
  {Schubert}}\ and\ \bibinfo {author} {\bibfnamefont {C.}~\bibnamefont
  {Gros}},\ }\href@noop {} {\bibfield  {journal} {\bibinfo  {journal}
  {Frontiers in computational neuroscience}\ }\textbf {\bibinfo {volume}
  {15}},\ \bibinfo {pages} {12} (\bibinfo {year} {2021})}\BibitemShut {NoStop}%
\bibitem [{\citenamefont {Shen}\ \emph {et~al.}(2017)\citenamefont {Shen},
  \citenamefont {Harris}, \citenamefont {Skirlo}, \citenamefont {Prabhu},
  \citenamefont {Baehr-Jones}, \citenamefont {Hochberg}, \citenamefont {Sun},
  \citenamefont {Zhao}, \citenamefont {Larochelle}, \citenamefont {Englund}
  \emph {et~al.}}]{shen2017deep}%
  \BibitemOpen
  \bibfield  {author} {\bibinfo {author} {\bibfnamefont {Y.}~\bibnamefont
  {Shen}}, \bibinfo {author} {\bibfnamefont {N.~C.}\ \bibnamefont {Harris}},
  \bibinfo {author} {\bibfnamefont {S.}~\bibnamefont {Skirlo}}, \bibinfo
  {author} {\bibfnamefont {M.}~\bibnamefont {Prabhu}}, \bibinfo {author}
  {\bibfnamefont {T.}~\bibnamefont {Baehr-Jones}}, \bibinfo {author}
  {\bibfnamefont {M.}~\bibnamefont {Hochberg}}, \bibinfo {author}
  {\bibfnamefont {X.}~\bibnamefont {Sun}}, \bibinfo {author} {\bibfnamefont
  {S.}~\bibnamefont {Zhao}}, \bibinfo {author} {\bibfnamefont {H.}~\bibnamefont
  {Larochelle}}, \bibinfo {author} {\bibfnamefont {D.}~\bibnamefont {Englund}},
   \emph {et~al.},\ }\href@noop {} {\bibfield  {journal} {\bibinfo  {journal}
  {Nature Photonics}\ }\textbf {\bibinfo {volume} {11}},\ \bibinfo {pages}
  {441} (\bibinfo {year} {2017})}\BibitemShut {NoStop}%
\bibitem [{\citenamefont {Inubushi}\ and\ \citenamefont
  {Yoshimura}(2017)}]{inubushi2017reservoir}%
  \BibitemOpen
  \bibfield  {author} {\bibinfo {author} {\bibfnamefont {M.}~\bibnamefont
  {Inubushi}}\ and\ \bibinfo {author} {\bibfnamefont {K.}~\bibnamefont
  {Yoshimura}},\ }\href@noop {} {\bibfield  {journal} {\bibinfo  {journal}
  {Scientific reports}\ }\textbf {\bibinfo {volume} {7}},\ \bibinfo {pages} {1}
  (\bibinfo {year} {2017})}\BibitemShut {NoStop}%
\end{thebibliography}%

\end{document}